\documentclass[conference]{IEEEtran}
\IEEEoverridecommandlockouts
\usepackage{cite}
\usepackage{amsmath,amssymb,amsfonts}
\usepackage{algorithmic}
\usepackage{graphicx}
\usepackage{textcomp}
\usepackage{xcolor}
\usepackage{multicol}
\usepackage{multirow}
\usepackage{tabularx}
\usepackage{booktabs}

\def\BibTeX{{\rm B\kern-.05em{\sc i\kern-.025em b}\kern-.08em
    T\kern-.1667em\lower.7ex\hbox{E}\kern-.125emX}}
\begin{document}

\title{UAV Base Station Location Optimization for Next Generation Wireless Networks: Overview and Future Research Directions}

\author{
\IEEEauthorblockN{Cihan Tugrul Cicek}
\IEEEauthorblockA{Department of Industrial Eng. \\
and Operations Research \\
University of California, Berkeley\\
Berkeley, California, USA \\
ctcicek@berkeley.edu}

\and

\IEEEauthorblockN{Hakan Gultekin}
\IEEEauthorblockA{Department of Mechanical \\
and Industrial Engineering \\
Sultan Qaboos University\\
Muscat, Oman \\
hgultekin@squ.edu.om}

\and

\IEEEauthorblockN{Bulent Tavli}
\IEEEauthorblockA{Department of Electrical\\
and Electronics Engineering \\
TOBB Univ. of Econ.\&Tech.\\
Ankara, Turkey \\
btavli@etu.edu.tr}

\and

\IEEEauthorblockN{Halim Yanikomeroglu}
\IEEEauthorblockA{Department of Systems\\
and Computer Engineering \\
Carleton University\\
Ottawa, Canada \\
halim@sce.carleton.ca}

}

\maketitle

\begin{abstract}
Unmanned aerial vehicles mounted base stations (UAV-BSs) are expected to become one of the significant components of the Next Generation Wireless Networks (NGWNs). Rapid deployment, mobility, higher chances of unobstructed propagation path, and flexibility features of UAV-BSs have attracted significant attention. Despite, potentially, high gains brought by UAV-BSs in NGWNs, many challenges are also introduced by them. Optimal location assignment to UAV-BSs, arguably, is the most widely investigated problem in the literature on UAV-BSs in NGWNs. This paper presents a comprehensive survey of the literature on the location optimization of UAV-BSs in NGWNs. A generic optimization framework through a universal Mixed Integer Non-Linear Programming (MINLP) formulation is constructed and the specifications of its constituents are elaborated. The generic problem is classified into a novel taxonomy. Due to the highly challenging nature of the optimization problem a range of solutions are adopted in the literature which are also covered under the aforementioned classification. Furthermore, future research directions on UAV-BS location optimization in 5G and beyond non-terrestrial aerial communication systems are discussed.
\end{abstract}


\section{Introduction}
Aerial base stations, which are also known as Unmanned Aerial Vehicle Base Stations (UAV-BSs) or drone-BSs have attracted significant interest in the past few years. Their rapid deployment, flexible relocation, and higher chances of experiencing Line-of-Sight (LoS) propagation path features have been perceived as promising opportunities to provide service in currently difficult to address service provisioning scenarios like short duration extremely crowded gatherings. Dynamically and selectively increasing the capacity of the network and improving the agility are also considered among the essential capabilities of beyond 5G cellular networks \cite{Bor-Yaliniz2016a}. Moreover, the coverage advantage of UAV-BSs over the terrestrial BSs' due to their higher operational altitudes highly likely to improve the performance of heterogeneous networks served by both the legacy terrestrial BSs and UAV-BSs collectively \cite{Chandrasekharan2016}. We use Flying Base Station (FBS) as the notation to denote all types of base stations that are either mounted on drones/UAVs or on platforms that have the ability to fly or float while serving as a wireless base station.

\begin{figure}[!b]
\resizebox{0.48\textwidth}{!}{\includegraphics{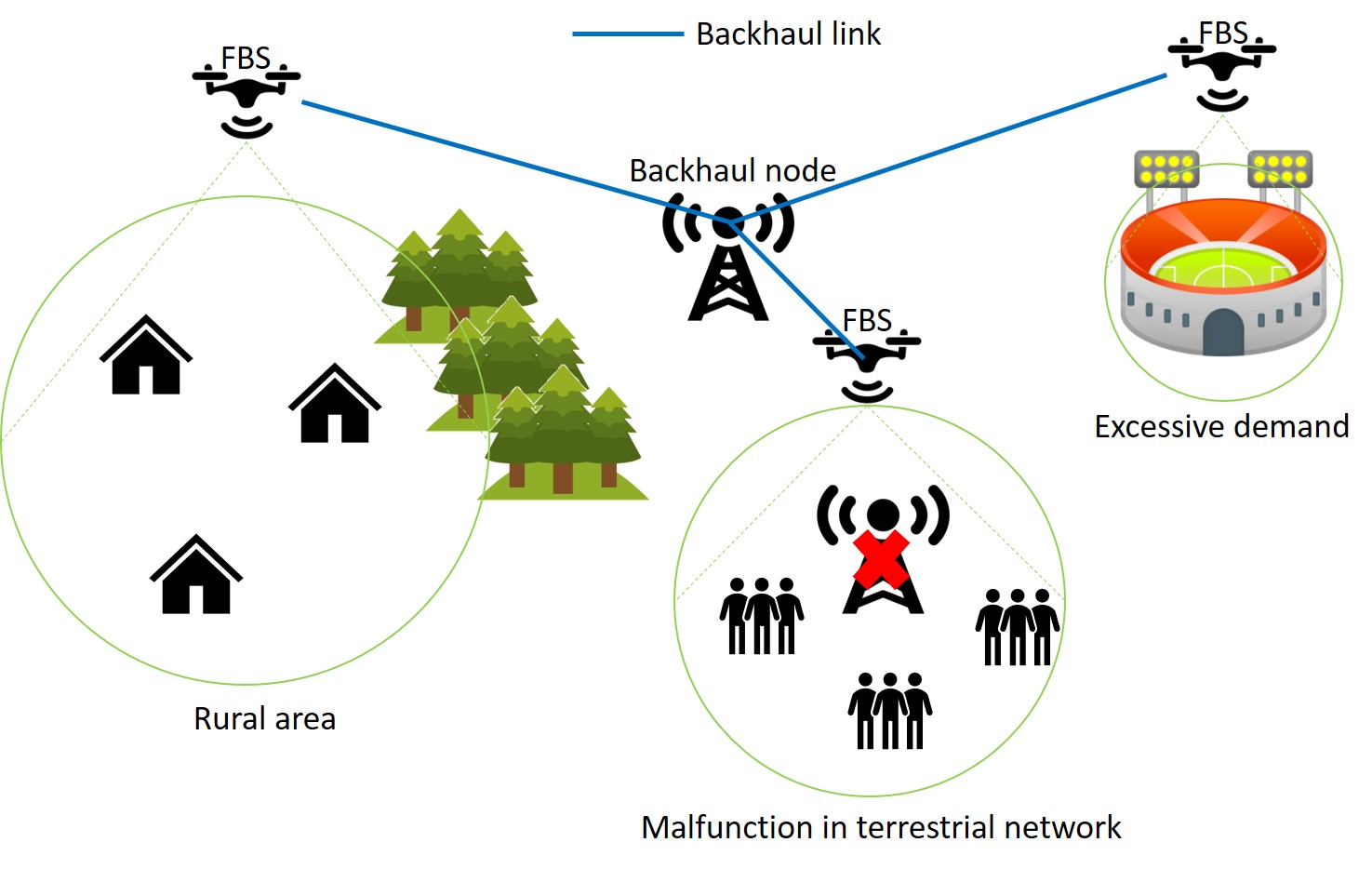}}
\caption{Illustration of use cases for FBSs.}
\label{fig1}
\end{figure}

Figure~\ref{fig1} illustrates a number of these that includes improving the Quality of Service (QoS) in sites where unexpected demand occurs (e.g., crowded sports events) \cite{Kalantari2017a}. As the deployments of terrestrial wireless networks are generally planned based on long-term traffic behaviour, matching the capacity with the demand at all times is not possible. This behaviour causes to either under-utilization of the resources or excess of capacity in the network. Therefore, a significant amount of resources are left idle in certain sites while fluctuating demand cannot be satisfied in other sites. To avoid such shortcomings of the existing wireless networks, FBSs are expected to help moving the excess capacity in the network towards where the demand occurs so that the network resources are utilized efficiently and the QoS improves significantly \cite{Kalantari2017b}. For example, it is proposed to offer limited incentives such as price reduction or high data rate to users with unsatisfactory coverage to move towards the better coverage regions served by FBSs \cite{Bor-Yaliniz2018}.

\begin{figure}[!h]
\resizebox{0.48\textwidth}{!}{\includegraphics{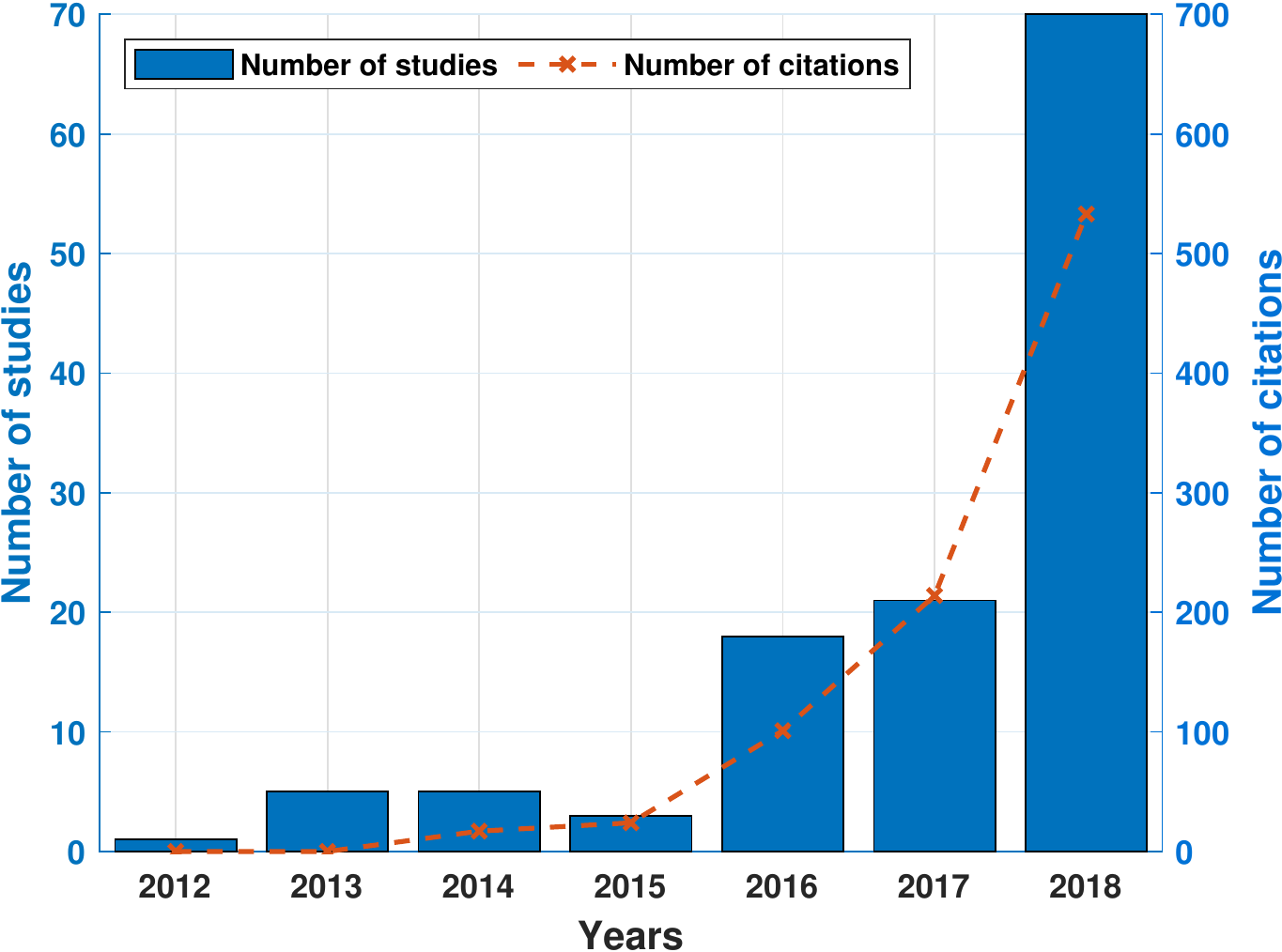}}
\caption{Number of studies and citations related to FBSs through the years.}
\label{fig2}
\end{figure}

The number of academic studies on the utilization of FBSs in NGWNs is increasing rapidly as well as the number of citations they receive as illustrated in Figure~\ref{fig2}. The data is compiled from the SCOPUS database by using search phrases ``aerial base station,'' ``UAV base station,'' ``drone base station,'' and ``flying base station''. Total number of papers that we encountered is 124 (earliest of which was published in 2012).

Despite the rapid built-up of the literature on FBSs for NGWNs in recent years, location optimization problem for FBSs have not been systematically surveyed to the best of our knowledge. It is worth mentioning that there are several survey/overview papers on FBSs \cite{Zeng2016,Gupta2016,Cao2018}, however these studies do not cover the FBS location optimization problems in detail. The FBS location optimization problems often include non-convexity in nature together with binary and continuous variables and are difficult to solve by either using existing solution methods or commercial solvers. Therefore, significant efforts have been put into devising novel approaches to address such complex problems. In this study, we create a novel and a detailed taxonomy of FBS location optimization problems and associated solution approaches. Within this taxonomy we surveyed and analyzed the literature to provide a concise digest of the current state-of-the-art which we believe will foster future research on FBS location optimization.

\section{Basic problem definition}
\label{sec:baseproblem}

FBSLP aims to determine the 3D locations of FBSs to improve certain sets of QoS objectives while satisfying predetermined sets of Service Level Requirements (SLRs) of the users and capacity restrictions of the FBSs. Note that the term ``user'' refers to any device that requests a certain amount of data transmitted through a communication channel. Let $J=\{1,\ldots ,n\}$ be the set of the FBSs, $I=\{1,\ldots ,m\}$ be the set of users, and $T=\{1,\ldots ,r\}$ be the set of periods. Let $X\in \mathbb{R}^{n\times 3}$ and $Y \in \mathbb{R}^{m\times 3}$ be the matrices that represent the coordinates of the FBSs (decision variables) and the users (parameters), respectively. It is obvious that the third column value of each row of $Y$ is equal to 0, since the users are usually assumed to be located on the ground level. Let $f_{it}:\mathbb{R}^3\times\mathbb{R}^3 \rightarrow \mathbb{R}$ be the function that defines the utility incurred when user $i$ is served, $g_{it}^k:\mathbb{R}^3\times\mathbb{R}^3 \rightarrow \mathbb{R}$ be the actual SLR level of user $i$ related to $k^{th}$ SLR, and $h_{jt}^l:\mathbb{R}^3\times\mathbb{R}^3 \rightarrow \mathbb{R}$ be the $l^{th}$ actual capacity level of FBS $j$ in period $t$. Note that, these functions are, typically, non-convex. Moreover, let $w_{j,t,t-1}:\mathbb{R}^3\times\mathbb{R}^3 \rightarrow \mathbb{R}$ be the function that governs the trajectory of the FBS $j$ in consecutive periods.

Assume that there exist $p$ different SLR levels for the users and $q$ different capacity levels for the FBSs. Let $v=\{1,\ldots,\eta\}$ be SLR-capacity pairs that directly associate with each other and $\eta \leq \min\{p,q\}$. FBSLP can be formulated as a Mixed Integer Non-Linear Programming (MINLP) formulation, which is presented in (P):

\begin{equation}
    \textrm{(P): } \underset{X}{\max} \sum_{i=1}^m\sum_{j=1}^n\sum_{t=1}^r u_{ijt}f_{it}(y_{it},x_{jt}) \nonumber
\label{eqn:obj}
\end{equation}
\begin{center}
   subject to
\end{center}
\begin{equation}
    s_{it}^k \leq \sum_{j=1}^ng_{it}^k(y_{it},x_{jt})u_{ijt},\forall i\in I,\forall t \in T,k=1,\ldots ,p
\label{eqn:user}
\end{equation}
\begin{equation}
    \sum_{j=1}^nu_{ijt} \leq 1,\forall i\in I,\forall t \in T
\label{eqn:userfbs}
\end{equation}
\begin{equation}
    \sum_{v=1}^\eta \sum_{i=1}^m g_{it}^v(y_{it},x_{jt})u_{ijt} \leq h_{jt}^v(x_{jt})z_{jt},\forall j\in J,\forall t \in T
\label{eqn:fbs}
\end{equation}
\begin{equation}
    w_{j,t,t-1}(x_{jt},x_{jt-1}) \leq C_{jt},\forall j\in J,t=2,\ldots ,r
\label{eqn:period}
\end{equation}
\begin{equation}
\begin{aligned}
    u_{ijt} \in \{ 0,1\},\forall i\in I,\forall j\in J, \forall t \in T \\
    z_{jt} \in \{ 0,1\},\forall j\in J, \forall t \in T \\
    D_{lo} \leq X \leq D_{up}.
\label{eqn:sign}
\end{aligned}
\end{equation}

In this formulation, $s_{it}^k$ is the $k^{th}$ SLR level requested by user $i$ in period $t$. $C_{jt}$ is the regulation parameter for FBS $j$ in period $t$ which is, typically, considered as the maximum allowed distance or velocity for an FBS in consecutive periods, since extreme changes in the distance or the velocity can deteriorate the performance of the FBSs in terms of Energy Consumption (EC) or movement capability. $u_{ijt}$ is the binary variable that is equal to 1 if the actual SLR level provided to user $i$ from FBS $j$ satisfies the requested level in period $t$. $z_{jt}$ is also a binary variable that is equal to 1 if FBS $j$ is activated in period $t$. $D_{lo}$ and $D_{up}$ are the vectors that define the boundaries of the region examined. Note that $y_{it}$ and $x_{jt}$ correspond to $i^{th}$ row of $Y$ and $j^{th}$ row of $X$ in period $t$, respectively. The objective function maximizes the total utility gained from all served users in a finite time horizon. Constraint set \eqref{eqn:user} ensures that only the users whose demands are satisfied can be served. Constraint set \eqref{eqn:userfbs} states that the users can be served by only one FBS. Constraint set \eqref{eqn:fbs} guarantees that capacities of the FBSs cannot be exceeded. Here, only the SLRs in $v$ is included in the left-hand side of the inequalities. Constraint set \eqref{eqn:period} prevents the FBSs to change their location in consecutive periods in a way that violates their movement capability (e.g., velocity). Constraint set \eqref{eqn:sign} defines the types and restrictions of the variables.

Utility, SLR, and capacity definitions in the FBSLP vary for different problem environments. The utility is usually defined as profit, sum-rate, or coverage. The SLR is defined as minimum threshold value of QoS examined (i.e., data rate). The capacity is related to backhaul capacity, Transmit Power (TP), or EC. In some cases, the objective may be \textit{minsum} type such as minimizing the total EC or the total mission completion time. This conversion can easily be realized by changing the objective function to $\min \sum_{j \in J}\sum_{t \in T}e_{jt}z_{jt}$, where $e_{jt}:\mathbb{R}^3\times\mathbb{R}^3 \rightarrow \mathbb{R}$ is the cost function of FBS $j$ in period $t$.

One of the main differences of FBSLP from the existing optimization problems in NGWNs is the air-to-ground (A2G) channel model that is generally used in $g_{it}^k$. Although several approaches have been proposed in the literature, the model by \cite{Al-Hourani2014} is the most cited one. In this model, the users are assumed to belong to one of the two propagation groups. One group consists of the users who have LoS with the FBSs, and the other group consists of the users who do not have LoS, however, such users can also communicate with the FBSs due to the mechanisms of electromagnetic wave propagation ($\rm e.g.$, reflection and diffraction) provided that the Signal-to-Noise Ratio (SNR) is above a certain threshold. Adopting this approach, we use the following notation to determine the SNR between user $i$ and FBS $j$ in period $t$, ${\textrm {SNR}}_{ijt}$:
\begin{itemize}
  \item $G_{jt}$: TP of FBS $j$ in period $t$ in $\rm dBm$,
  \item $f_c$: carrier frequency in $\rm Hz$,
  \item $c$: speed of light in $\rm m/s$,
  \item $d_{ijt}$: distance between user $i$ and FBS $j$ in period $t$ in meter,
  \item $b_{ijt}$: bandwidth allocated to user $i$ by FBS $j$ in period $t$ in $\rm Hz$,
  \item $\theta_{ijt}$: elevation angle between user $i$ and FBS $j$ in period $t$ in degrees,
  \item $\eta, \ \alpha ,\ \beta ,\ \varphi_{\rm LoS},\ \varphi_{\rm NLoS}$: parameters varying according to the environment.
\end{itemize}
Then, ${\textrm {SNR}}_{ijt}$ can be found by \eqref{eqn:snr}:

\begin{equation}
\begin{split}
    {\textrm {SNR}}_{ijt}=A - 10\eta{\rm log}_{10}(d_{ijt}) \\
    +\frac{B}{1+{\alpha}e^{-\beta(\theta_{ijt}-\alpha)}}-10{\rm log}_{10}(b_{ijt})
\end{split}
\label{eqn:snr}
\end{equation}

\noindent where $A=G_{jt}-10\eta{\rm log}(\frac{4\pi f_c}{c})-\varphi_{\rm NLoS}$ and $B=\varphi_{\rm LoS}-\varphi_{\rm NLoS}$. It has been shown that when $b_{ijt}$ is fixed, if a local maxima exists in \eqref{eqn:snr}, then it is the only local maxima for a given SNR value~\cite{Bor-Yaliniz2016b}. Although this is a strong property to use, manipulating \eqref{eqn:snr} is still complicated since it is neither a convex nor a concave function. In fact, SNR increases with respect to the increasing altitude of the FBS due to the LoS advantage. However, impact of the altitude has a negative effect on the SNR as the distance dominates \eqref{eqn:snr} after this threshold.

\section{Classification}
\label{sec:classification}
Problem (P) includes a non-convex objective function and constraints with binary and continuous variables which renders finding an unambiguous solution a challenging task. In fact, the FBSLP belongs to the NP-hard problem class as, with a number of assumptions, it can be reduced to the well-studied set covering or maximal covering location problems which are shown to be NP-hard \cite{Farahani2012}. Therefore, various approaches such as decomposition and heuristic algorithms have been proposed to solve the FBS problem instances in reasonable time. The solution approaches have been commonly focused on two major sets of the problems, $X$ (coordinates of the FBSs) and $Y$ (coordinates of the users). The solution approaches differentiate from each other by the assumptions of mobile FBSs only, mobile users only, or both mobile users and FBSs.

\begin{table*}
\caption{Comparison of the papers}
\begin{tabularx}{\textwidth}{p{.55cm} p{.3cm} p{1.8cm} p{2.2cm} p{3cm} X  p{2.8cm}}
\toprule
Paper & FBS & Utility & Cost & SLR & Capacity & Solution  \\
\midrule
\multicolumn{7}{l}{SFBSLP} \\
\midrule

\cite{Bor-Yaliniz2016a} & S & User number & - & SNR & Backhaul capacity & Exact   \\

\cite{Kalantari2017a} & S & User number & - & SNR & Backhaul capacity & ENUM  \\

\cite{Kalantari2017b} & M & Sum-rate & - & - & Backhaul capacity, Cover all users, No overlap among the FBSs & PSH\\

\cite{Bor-Yaliniz2018} & S & Combined function* & - & Pathloss & - & Exact, PSH  \\

\cite{Bor-Yaliniz2016b} & S & User number & - & SNR & - & Exact  \\

\cite{Cicek2018} & S & Profit & - & Rate & Backhaul capacity & PSH  \\

\cite{Alzenad2017} & S & User number & - & SNR & - & Exact (MOSEK)  \\

\cite{Mozaffari2015} & M & Coverage & TP & SNR & - & Exact   \\

\cite{Mozaffari2016a} & M & Coverage & - & SNR & - & PSH  \\

\cite{Merwaday2015} & M & SE & - & SE & - & ENUM  \\

\cite{Lyu2017a} & M & - & FBS number & - & Cover all users & PSH  \\

\cite{Merwaday2016} & M & SE & - & - & - & GA**  \\

\cite{ChenJ2017} & S & Sum-rate & TP & Rate & Maximum rate & Exact (single user case) \\

\cite{Sharma2016a} & M & - & Latency & - & Discrete locations & LA  \\

\cite{Shakhatreh2017a} & S & - & TP & Rate & Cover all users & PSO  \\

\cite{Shakhatreh2017b} & S & - & TP & Rate & Cover all users & GDA  \\

\cite{Rupasinghe2016} & M & SNR & - & Interference & - & PSH  \\

\cite{Kalantari2016} & M & - & FBS number & SNR & Cover all users & Exact  \\

\cite{Alzenad2018} & S & User number & - & SNR & - & ENUM, PSH  \\

\cite{Koyuncu2018} & M & - & TP & Rate & - & Exact**  \\

\cite{Sun2018} & S & SE & - & - & - & PSH  \\

\cite{Lagum2018} & M & - & Spatial irregularity & - & Exact number of the FBSs to be selected & PSH  \\

\midrule
\multicolumn{7}{l}{SMFBSLP} \\
\midrule

\cite{Zeng2017} & S & Sum-rate & TP & - & Maximum time per period, Maximum velocity per period & PSH**  \\

\cite{Mozaffari2016c} & M & - & TP & Rate & - & PSH  \\

\cite{Mozaffari2016d} & M & - & TP & SNR & Minimum number of users & PSH  \\

\cite{Mozaffari2017} & M & - & TP & SNR & - & PSH  \\

\cite{Wu2018a} & M & Minimum rate & - & One-on-one assignment & Maximum distance per period & PSH  \\

\cite{Zhang2017} & S & Secrecy & - & Rate & Maximum velocity per period & PSH  \\

\cite{Xie2018} & S & Minimum rate & - & Energy neutrality & Maximum TP & PSH  \\

\cite{Zeng2018} & M & - & ST & SNR & Maximum velocity per period & PSH \\

\cite{Lyu2017b} & M & Minimum rate & - & Outage & Maximum velocity per period & PSH  \\

\cite{Cheng2018} & S & Sum-rate & - & Rate &  & PSH**  \\

\cite{Alsharoa2017} & M & - & EC & - & Maximum consumed energy per period, Maximum stored energy per period & Exact (CPLEX) \\

\cite{Ghazzai2017} & S & - & EC & Rate & - & PSO, PSH  \\

\cite{Sun2017} & S & - & Latency & - & Capacity on energy & PSH  \\

\cite{Wu2018b} & S & Average rate & - & - & Maximum velocity per period, Maximum average TP & PSH  \\

\cite{YangD2018} & S & Sum-rate & EC & Rate & Maximum time to fly per period, Maximum EC per period & Exact**  \\

\cite{Challita2018} & M & - & Combined cost function*** & Demand & Capacity on TP, Receive from at least one ground BS, Cover all users & LA  \\

\cite{Wu2017} & S & Minimum rate & - & - & Maximum distance per period & PSH**  \\

\cite{Mozaffari2018} & M & - & ST & - & Maximum velocity per period, Minimum distance between two adjacent FBS in the array & PSH  \\

\cite{Bulut2018} & M & - & ST & - & Maximum disconnectivity time & DP**  \\

\midrule
\multicolumn{7}{l}{DFBSLP} \\
\midrule

\cite{ChenM2017} & M & - & TP & Combined QoE & - & LA  \\

\cite{YangP2017} & M & Coverage & - & SNR & Backhaul capacity & PSH**  \\

\cite{Fotouhi2017} & S & SE & - & - & Maximum velocity per period, At most 1 user is served by FBS per period & PSH**  \\

\cite{Fotouhi2017} & M & - & TP & Rate & Maximum movement per period & Exact  \\

\cite{Ghanavi2018} & S & Sum-rate & - & - & Cover all users & LA  \\
\bottomrule
\multicolumn{7}{l}{S: Single,  M: Multiple} \\
\multicolumn{7}{l}{* Utility function includes two components: User number + Profit} \\
\multicolumn{7}{l}{** The altitude of the FBSs are assumed to be fixed.} \\
\multicolumn{7}{l}{*** Cost function includes three components: Total interference + time to complete the tasks + latency.}
\end{tabularx}
\label{table}
\end{table*}

We classified literature on FBS location optimization under three major categories based on how $X$ and $Y$ sets are treated. The first category, namely static FBSLP (SFBSLP), has focused on solving the problem for a snapshot of the time where both users and the FBSs are assumed to be not moving. The second category, namely semi-dynamic FBSLP (SMFBSLP), relaxes this assumption in $J$ and allows the FBSs to move to improve the performance of the network in a finite time horizon. The last group, namely dynamic FBSLP (DFBSLP), relaxes the stationary assumption in $I$ as well, where both the users and the FBSs are allowed to move. Note that the set $T$ becomes redundant in SFBSLP and can be removed from (P) together with constraints~\eqref{eqn:period} and all subscripts in other parameters and variables. In SMFBSLP and DFBSLP, users are generally assumed to move following a predefined random pattern in periods, hence, $Y$ is explicitly provided for each period. Moreover, all problems can be divided into two more cases depending on locating a single FBS ($n=1$) or multiple FBSs. The latter one is more complex since one more dimension is added to the search space.

One of the main components of the FBSLP is the utility function. Recall that $f_{it}$ defines the amount of utility gained when user $i$ is served in period $t$. In \textit{maxsum} problems, the utility may refer to an indicator that represents whether the user is covered/served or not (alternatively utility can be the level of the service like the rate provided to the user). Other utility functions refer to the Spectral Efficiency (SE) or the SNR of the user. SE and SNR can also be considered as SLR levels, since the value under a minimum threshold of the SE or the SNR implies an unreliable service. In \textit{minsum} problems, the utility is replaced with a cost function and is associated with the FBSs rather than the users. The costs refer to TP, EC, or latency. In a limited number of problems, in which the FBSs are used to dispatch the data packages, the cost refers to the service time (ST).

Other components of the FBSLP are the SLR and capacity functions. Most of the studies discussed in this paper do not consider the SLR and the capacity in the same problem. Instead, the problems are subject to either up to two SLR requirements for the users or up to three capacity restrictions for the FBSs. The SLR functions may refer to SE, rate, SNR, or demand for data packets, while the capacity functions may refer to the number of active FBSs, the total time that the FBSs can hover, and a number of QoS constraints such as the backhaul capacity, total TP, or the minimum utilization ratio. For the dynamic problems, there are usually additional constraints that require the FBSs to return back to their starting locations, which may be the charging station or to another location to be dispatched.

Table~\ref{table} presents the FBS location optimization literature categorized according to the classification methodology we created and explained in this section. Due to the space limitations, we cannot include all the papers in the literature on FBS for NGWNs, however, we compiled a large subset of the literature with the highest relevance and highest number of citations. The features we employed to classify the solution approaches are as follows:
\begin{itemize}
  \item \textbf{Exact: }The approach guarantees to find the global optimum or is able to provide the worst case deviation from the optimal if stopped early,
  \item \textbf{Given heuristic names: }The approaches are well-known Dynamic Programming (DP) or meta-heuristics such as Particle Swarm Optimization algorithm (PSO), Genetic Algorithm (GA), or Gradient Algorithm (GDA),
  \item \textbf{Learning algorithm (LA): }The approach benefits a learning procedure such as reinforcement learning,
  \item \textbf{Enumeration (ENUM): }The approach uses an exhaustive search technique to find the best solution,
  \item \textbf{Problem specific heuristic (PSH): }An ad hoc approach tailored according to the problem properties.
\end{itemize}

One of the important trends in FBS literature is the shift of the focus from static problems to the dynamic ones. For example, $74\%$ of the papers in 2018 were on dynamic problems which was only $25\%$ in 2015. Note that, the assumption of static users do not reflect the real-world accurately. However, SFBSLP is still likely to receive further attention in future because of the convenience of analyzing static environments.

Solving the optimization problems on FBSLP is a challenging task. Generally, it is not possible to adopt off-the-shelf approaches, therefore, most of the researchers created problem specific solution approaches. For example, $56\%$ of the studies proposed a new heuristic (PSH) and additional $10\%$ have adopted a meta-heuristic with specific adjustment to parameters of the algorithms.

SLR is usually considered as the data rate requested by the users or SNR value. The most adopted capacity constraints in the literature are the distance that the FBSs move or the velocity that the FBSs can hover in a period and the coverage threshold. Sum-rate and coverage are the most frequently employed objectives to be maximized, while energy related objectives such as the total TP and the total EC are the most used cost functions.

\section{Conclusion and future research}
\label{sec:conclusion}

In this paper, we present a concise overview of the optimization approaches to solve the location problem of FBSs for NGWNs which are envisioned to be the integral constituents of the future networks (i.e., 5G and beyond). Not only do we overview the solution methodologies in the literature, but also we give the general form of the mathematical formulation of the FBS location problems. Literature on FBSs has substantially grown in recent years due to their unique capabilities like rapid deployment and flexibility. However, FBSs have brought significant challenges to be addressed to take complete advantage of the opportunities that arise. One of the important future research avenues is to ensure reliability of the services provided by FBSs. The users expect ubiquitous and reliable connections while moving on the ground, therefore, reliable backhaul links should be integrated to the heterogeneous networks where the FBSs are utilized. Another important research area is to address the problem of limited battery-dependent operational service span of the FBSs. Indeed, many studies have shown that the service time of the FBSs is limited and novel techniques should be adopted to boost the energy efficiency. Otherwise, it is unlikely that the envisioned application scenarios of FBSs will ever receive widespread adoption. Developing versatile and faster solution algorithms for the FBS location problems is also a promising future research topic. For example, learning based optimization algorithms have a potential to handle unexpected changes in the problem environment in comparison to specific algorithms developed to solve predefined problem structures.

\bibliographystyle{IEEEtran}
\bibliography{conference_041818}

\begin{thebibliography}{10}
\providecommand{\url}[1]{#1}
\csname url@samestyle\endcsname
\providecommand{\newblock}{\relax}
\providecommand{\bibinfo}[2]{#2}
\providecommand{\BIBentrySTDinterwordspacing}{\spaceskip=0pt\relax}
\providecommand{\BIBentryALTinterwordstretchfactor}{4}
\providecommand{\BIBentryALTinterwordspacing}{\spaceskip=\fontdimen2\font plus
\BIBentryALTinterwordstretchfactor\fontdimen3\font minus
  \fontdimen4\font\relax}
\providecommand{\BIBforeignlanguage}[2]{{%
\expandafter\ifx\csname l@#1\endcsname\relax
\typeout{** WARNING: IEEEtran.bst: No hyphenation pattern has been}%
\typeout{** loaded for the language `#1'. Using the pattern for}%
\typeout{** the default language instead.}%
\else
\language=\csname l@#1\endcsname
\fi
#2}}
\providecommand{\BIBdecl}{\relax}
\BIBdecl

\bibitem{Bor-Yaliniz2016a}
I.~Bor-Yaliniz and H.~Yanikomeroglu, ``The new frontier in {RAN} heterogeneity:
  {Multi}-tier drone-cells,'' \emph{IEEE Communications Magazine}, vol.~54,
  no.~11, pp. 48--55, November 2016.

\bibitem{Chandrasekharan2016}
S.~Chandrasekharan, K.~Gomez, A.~Al-Hourani, S.~Kandeepan, T.~Rasheed,
  L.~Goratti, L.~Reynaud, D.~Grace, I.~Bucaille, T.~Wirth, and S.~Allsopp,
  ``Designing and implementing future aerial communication networks,''
  \emph{IEEE Communications Magazine}, vol.~54, no.~5, pp. 26--34, May 2016.

\bibitem{Kalantari2017a}
E.~Kalantari, M.~Z. Shakir, H.~Yanikomeroglu, and A.~Yongacoglu,
  ``Backhaul-aware robust {3D} drone placement in {5G+} wireless networks,'' in
  \emph{2017 IEEE International Conference on Communications Workshops (ICC
  Workshops)}, May 2017, pp. 109--114.

\bibitem{Kalantari2017b}
E.~Kalantari, I.~Bor-Yaliniz, A.~Yongacoglu, and H.~Yanikomeroglu, ``User
  association and bandwidth allocation for terrestrial and aerial base stations
  with backhaul considerations,'' in \emph{2017 IEEE 28th Annual International
  Symposium on Personal, Indoor, and Mobile Radio Communications (PIMRC)}, Oct
  2017, pp. 1--6.

\bibitem{Bor-Yaliniz2018}
R.~I. Bor-Yaliniz, A.~El-Keyi, and H.~Yanikomeroglu, ``Spatial configuration of
  agile wireless networks with drone-{BS}s and user-in-the-loop,'' \emph{{\rm
  to appear in }IEEE Transactions on Wireless Communications}, pp. 1--16, 2018.

\bibitem{Zeng2016}
Y.~Zeng, R.~Zhang, and T.~J. Lim, ``Wireless communications with unmanned
  aerial vehicles: Opportunities and challenges,'' \emph{IEEE Communications
  Magazine}, vol.~54, no.~5, pp. 36--42, May 2016.

\bibitem{Gupta2016}
L.~Gupta, R.~Jain, and G.~Vaszkun, ``Survey of important issues in {UAV}
  communication networks,'' \emph{IEEE Communications Surveys Tutorials},
  vol.~18, no.~2, pp. 1123--1152, Secondquarter 2016.

\bibitem{Cao2018}
X.~Cao, P.~Yang, M.~Alzenad, X.~Xi, D.~Wu, and H.~Yanikomeroglu, ``Airborne
  communication networks: A survey,'' \emph{{\rm to appear in }IEEE Journal on
  Selected Areas in Communications}, pp. 1--20, 2018.

\bibitem{Al-Hourani2014}
A.~Al-Hourani, S.~Kandeepan, and S.~Lardner, ``Optimal {LAP} altitude for
  maximum coverage,'' \emph{IEEE Wireless Communications Letters}, vol.~3,
  no.~6, pp. 569--572, Dec. 2014.

\bibitem{Bor-Yaliniz2016b}
R.~I. Bor-Yaliniz, A.~El-Keyi, and H.~Yanikomeroglu, ``Efficient {3-D}
  placement of an aerial base station in next generation cellular networks,''
  in \emph{2016 IEEE International Conference on Communications (ICC)}, May
  2016, pp. 1--5.

\bibitem{Farahani2012}
R.~Z. Farahani, N.~Asgari, N.~Heidari, M.~Hosseininia, and M.~Goh, ``Covering
  problems in facility location: A review,'' \emph{Computers {\&} Industrial
  Engineering}, vol.~62, no.~1, pp. 368 -- 407, Feb 2012.

\bibitem{Cicek2018}
C.~{Tugrul Cicek}, T.~{Kutlu}, H.~{Gultekin}, B.~{Tavli}, and
  H.~{Yanikomeroglu}, ``{Backhaul-Aware Placement of a UAV-BS with Bandwidth
  Allocation for User-Centric Operation and Profit Maximization},'' \emph{ArXiv
  e-prints}, p. arXiv:1810.12395, Oct. 2018.

\bibitem{Alzenad2017}
M.~Alzenad, A.~El-Keyi, F.~Lagum, and H.~Yanikomeroglu, ``{3-D} placement of an
  unmanned aerial vehicle base station {(UAV-BS)} for energy-efficient maximal
  coverage,'' \emph{IEEE Wireless Communications Letters}, vol.~6, no.~4, pp.
  434--437, Aug 2017.

\bibitem{Mozaffari2015}
M.~Mozaffari, W.~Saad, M.~Bennis, and M.~Debbah, ``Drone small cells in the
  clouds: Design, deployment and performance analysis,'' in \emph{2015 IEEE
  Global Communications Conference (GLOBECOM)}, Dec 2015, pp. 1--6.

\bibitem{Mozaffari2016a}
------, ``Efficient deployment of multiple unmanned aerial vehicles for optimal
  wireless coverage,'' \emph{IEEE Communications Letters}, vol.~20, no.~8, pp.
  1647--1650, Aug 2016.

\bibitem{Merwaday2015}
A.~Merwaday and I.~Guvenc, ``{UAV} assisted heterogeneous networks for public
  safety communications,'' in \emph{2015 IEEE Wireless Communications and
  Networking Conference Workshops}, March 2015, pp. 329--334.

\bibitem{Lyu2017a}
J.~Lyu, Y.~Zeng, R.~Zhang, and T.~J. Lim, ``Placement optimization of
  {UAV}-mounted mobile base stations,'' \emph{IEEE Communications Letters},
  vol.~21, no.~3, pp. 604--607, March 2017.

\bibitem{Merwaday2016}
A.~Merwaday, A.~Tuncer, A.~Kumbhar, and I.~Guvenc, ``Improved throughput
  coverage in natural disasters: Unmanned aerial base stations for
  public-safety communications,'' \emph{IEEE Vehicular Technology Magazine},
  vol.~11, no.~4, pp. 53--60, Dec 2016.

\bibitem{ChenJ2017}
J.~Chen and D.~Gesbert, ``Optimal positioning of flying relays for wireless
  networks: A {LOS} map approach,'' in \emph{2017 IEEE International Conference
  on Communications (ICC)}, May 2017, pp. 1--6.

\bibitem{Sharma2016a}
V.~Sharma, R.~Sabatini, and S.~Ramasamy, ``{UAVs} assisted delay optimization
  in heterogeneous wireless networks,'' \emph{IEEE Communications Letters},
  vol.~20, no.~12, pp. 2526--2529, Dec 2016.

\bibitem{Shakhatreh2017a}
H.~Shakhatreh, A.~Khreishah, A.~Alsarhan, I.~Khalil, A.~Sawalmeh, and N.~S.
  Othman, ``Efficient {3D} placement of a {UAV} using particle swarm
  optimization,'' in \emph{2017 8th International Conference on Information and
  Communication Systems}, April 2017, pp. 258--263.

\bibitem{Shakhatreh2017b}
H.~Shakhatreh, A.~Khreishah, and B.~Ji, ``Providing wireless coverage to
  high-rise buildings using {UAVs},'' in \emph{2017 IEEE International
  Conference on Communications (ICC)}, May 2017, pp. 1--6.

\bibitem{Rupasinghe2016}
N.~Rupasinghe, A.~S. Ibrahim, and I.~Guvenc, ``Optimum hovering locations with
  angular domain user separation for cooperative {UAV} networks,'' in
  \emph{2016 IEEE Global Communications Conference (GLOBECOM)}, Dec 2016, pp.
  1--6.

\bibitem{Kalantari2016}
E.~Kalantari, H.~Yanikomeroglu, and A.~Yongacoglu, ``On the number and {3D}
  placement of drone base stations in wireless cellular networks,'' in
  \emph{2016 IEEE 84th Vehicular Technology Conference (VTC-Fall)}, Sept 2016,
  pp. 1--6.

\bibitem{Alzenad2018}
M.~Alzenad, A.~El-Keyi, and H.~Yanikomeroglu, ``{3-D} placement of an unmanned
  aerial vehicle base station for maximum coverage of users with different
  {QoS} requirements,'' \emph{IEEE Wireless Communications Letters}, vol.~7,
  no.~1, pp. 38--41, Feb 2018.

\bibitem{Koyuncu2018}
E.~Koyuncu, R.~Khodabakhsh, N.~Surya, and H.~Seferoglu, ``Deployment and
  trajectory optimization for {UAVs}: A quantization theory approach,'' in
  \emph{2018 IEEE Wireless Communications and Networking Conference}, April
  2018, pp. 1--6.

\bibitem{Sun2018}
X.~Sun and N.~Ansari, ``Jointly optimizing drone-mounted base station placement
  and user association in heterogeneous networks,'' in \emph{2018 IEEE
  International Conference on Communications (ICC)}, May 2018, pp. 1--6.

\bibitem{Lagum2018}
F.~Lagum, I.~Bor-Yaliniz, and H.~Yanikomeroglu, ``Strategic densification with
  {UAV-BSs} in cellular networks,'' \emph{IEEE Wireless Communications
  Letters}, vol.~7, no.~3, pp. 384--387, June 2018.

\bibitem{Zeng2017}
Y.~Zeng and R.~Zhang, ``Energy-efficient {UAV} communication with trajectory
  optimization,'' \emph{IEEE Transactions on Wireless Communications}, vol.~16,
  no.~6, pp. 3747--3760, June 2017.

\bibitem{Mozaffari2016c}
M.~Mozaffari, W.~Saad, M.~Bennis, and M.~Debbah, ``Optimal transport theory for
  power-efficient deployment of unmanned aerial vehicles,'' in \emph{2016 IEEE
  International Conference on Communications (ICC)}, May 2016, pp. 1--6.

\bibitem{Mozaffari2016d}
------, ``Mobile internet of things: Can {UAVs} provide an energy-efficient
  mobile architecture?'' in \emph{2016 IEEE Global Communications Conference
  (GLOBECOM)}, Dec 2016, pp. 1--6.

\bibitem{Mozaffari2017}
------, ``Mobile unmanned aerial vehicles {(UAVs)} for energy-efficient
  internet of things communications,'' \emph{IEEE Transactions on Wireless
  Communications}, vol.~16, no.~11, pp. 7574--7589, Nov 2017.

\bibitem{Wu2018a}
Q.~Wu, Y.~Zeng, and R.~Zhang, ``Joint trajectory and communication design for
  multi-{UAV} enabled wireless networks,'' \emph{IEEE Transactions on Wireless
  Communications}, vol.~17, no.~3, pp. 2109--2121, March 2018.

\bibitem{Zhang2017}
G.~Zhang, Q.~Wu, M.~Cui, and R.~Zhang, ``Securing {UAV} communications via
  trajectory optimization,'' in \emph{2017 IEEE Global Communications
  Conference (GLOBECOM)}, Dec 2017, pp. 1--6.

\bibitem{Xie2018}
L.~Xie, J.~Xu, and R.~Zhang, ``Throughput maximization for {UAV}-enabled
  wireless powered communication networks,'' in \emph{2018 IEEE 87th Vehicular
  Technology Conference (VTC-Spring)}, June 2018, pp. 1--7.

\bibitem{Zeng2018}
Y.~Zeng, X.~Xu, and R.~Zhang, ``Trajectory design for completion time
  minimization in {UAV}-enabled multicasting,'' \emph{IEEE Transactions on
  Wireless Communications}, vol.~17, no.~4, pp. 2233--2246, April 2018.

\bibitem{Lyu2017b}
J.~Lyu, Y.~Zeng, and R.~Zhang, ``Spectrum sharing and cyclical multiple access
  in {UAV}-aided cellular offloading,'' in \emph{2017 IEEE Global
  Communications Conference (GLOBECOM)}, Dec 2017, pp. 1--6.

\bibitem{Cheng2018}
F.~Cheng, S.~Zhang, Z.~Li, Y.~Chen, N.~Zhao, F.~R. Yu, and V.~C.~M. Leung,
  ``{UAV} trajectory optimization for data offloading at the edge of multiple
  cells,'' \emph{IEEE Transactions on Vehicular Technology}, vol.~67, no.~7,
  pp. 6732--6736, July 2018.

\bibitem{Alsharoa2017}
A.~Alsharoa, H.~Ghazzai, A.~Kadri, and A.~E. Kamal, ``Energy management in
  cellular {HetNets} assisted by solar powered drone small cells,'' in
  \emph{2017 IEEE Wireless Communications and Networking Conference}, March
  2017, pp. 1--6.

\bibitem{Ghazzai2017}
H.~Ghazzai, M.~B. Ghorbel, A.~Kadri, and M.~J. Hossain, ``Energy efficient {3D}
  positioning of micro unmanned aerial vehicles for underlay cognitive radio
  systems,'' in \emph{2017 IEEE International Conference on Communications
  (ICC)}, May 2017, pp. 1--6.

\bibitem{Sun2017}
X.~Sun and N.~Ansari, ``Latency aware drone base station placement in
  heterogeneous networks,'' in \emph{2017 IEEE Global Communications Conference
  (GLOBECOM)}, Dec 2017, pp. 1--6.

\bibitem{Wu2018b}
Y.~Wu, J.~Xu, L.~Qiu, and R.~Zhang, ``Capacity of {UAV}-enabled multicast
  channel:joint trajectory design and power allocation,'' in \emph{2018 IEEE
  International Conference on Communications(ICC)}, May 2018, pp. 1--7.

\bibitem{YangD2018}
D.~Yang, Q.~Wu, Y.~Zeng, and R.~Zhang, ``Energy tradeoff in ground-to-{UAV}
  communication via trajectory design,'' \emph{IEEE Transactions on Vehicular
  Technology}, vol.~67, no.~7, pp. 6721--6726, July 2018.

\bibitem{Challita2018}
U.~Challita, W.~Saad, and C.~Bettstetter, ``Deep reinforcement learning for
  interference-aware path planning of cellular-connected {UAVs},'' in
  \emph{2018 IEEE International Conference on Communications (ICC)}, May 2018,
  pp. 1--7.

\bibitem{Wu2017}
Q.~Wu, Y.~Zeng, and R.~Zhang, ``Joint trajectory and communication design for
  {UAV}-enabled multiple access,'' in \emph{2017 IEEE Global Communications
  Conference (GLOBECOM)}, Dec 2017, pp. 1--6.

\bibitem{Mozaffari2018}
M.~Mozaffari, W.~Saad, M.~Bennis, and M.~Debbah, ``Communications and control
  for wireless drone-based antenna array,'' \emph{IEEE Transactions on
  Communications}, pp. 1--1, 2018.

\bibitem{Bulut2018}
E.~Bulut and I.~Guevenc, ``Trajectory optimization for cellular-connected
  {UAVs} with disconnectivity constraint,'' in \emph{2018 IEEE International
  Conference on Communications Workshops (ICC Workshops)}, May 2018, pp. 1--6.

\bibitem{ChenM2017}
M.~Chen, M.~Mozaffari, W.~Saad, C.~Yin, M.~Debbah, and C.~S. Hong, ``Caching in
  the sky: Proactive deployment of cache-enabled unmanned aerial vehicles for
  optimized quality-of-experience,'' \emph{IEEE Journal on Selected Areas in
  Communications}, vol.~35, no.~5, pp. 1046--1061, May 2017.

\bibitem{YangP2017}
P.~Yang, X.~Cao, C.~Yin, Z.~Xiao, X.~Xi, and D.~Wu, ``Proactive drone-cell
  deployment: Overload relief for a cellular network under flash crowd
  traffic,'' \emph{IEEE Transactions on Intelligent Transportation Systems},
  vol.~18, no.~10, pp. 2877--2892, Oct 2017.

\bibitem{Fotouhi2017}
A.~Fotouhi, M.~Ding, and M.~Hassan, ``Dynamic base station repositioning to
  improve spectral efficiency of drone small cells,'' in \emph{2017 IEEE 18th
  International Symposium on A World of Wireless, Mobile and Multimedia
  Networks}, June 2017, pp. 1--9.

\bibitem{Ghanavi2018}
R.~Ghanavi, E.~Kalantari, M.~Sabbaghian, H.~Yanikomeroglu, and A.~Yongacoglu,
  ``Efficient {3D} aerial base station placement considering users mobility by
  reinforcement learning,'' in \emph{2018 IEEE Wireless Communications and
  Networking Conference}, April 2018, pp. 1--6.

\end{thebibliography}

\end{document}